\documentclass[twocolumn,preprintnumbers,superscriptaddress,nofootinbib,aps,floatfix]{revtex4-1}

\usepackage{graphicx}  
\usepackage{dcolumn}   
\usepackage{bm}        
\usepackage{amssymb,amsmath}
\usepackage{wasysym}
\usepackage{adjustbox,lipsum}
\usepackage{subfigure}

\def\beq{\begin{equation}}
\def\eeq{\end{equation}}
\usepackage[usenames,dvipsnames,svgnames]{xcolor}
\usepackage[dvipsnames]{xcolor}
\usepackage[colorlinks=true, allcolors=blue]{hyperref}
\hypersetup{colorlinks, citecolor=black, linkcolor=black, urlcolor=black}

\usepackage{textcomp}
\usepackage{amsmath}
\usepackage{amsfonts}
\usepackage{amssymb}
\usepackage{bbm}                
\usepackage{graphicx, rotating}
\usepackage{epstopdf}
\usepackage{epsfig}
\usepackage{latexsym}
\usepackage{graphicx}
\usepackage{bbold}
\usepackage{color}
\usepackage{amsmath,amssymb}
\usepackage{slashed}
\usepackage{cleveref}

\newcommand{\bea}{\begin{eqnarray}}
\newcommand{\eea}{\end{eqnarray}}
\newcommand{\eq}[1]{Eq.~(\ref{#1})}

\newcommand{\nn}{\nonumber}

\def\stw{s_{\theta_W}}
\def\ctw{c_{\theta_W}}
\def\ttw{t_{\theta_W}}
\def\lra#1{\overset{\text{\scriptsize$\leftrightarrow$}}{#1}}

\preprint{IPPP/19/35}
\begin{document}

\title{Resolving the  tensor structure of the Higgs coupling to $Z$-bosons\\  via Higgs-strahlung}
\author{Shankha Banerjee, Rick S. Gupta, Joey Y. Reiness and Michael Spannowsky}
\affiliation{Institute for Particle Physics Phenomenology,
Durham University, South Road, Durham, DH1 3LE}

\date{\today}

\begin{abstract}
We propose differential observables for  $pp \to Z(\ell^+\ell^-)h(b\bar{b})$   that can be used to completely determine the tensor structure of the $hZZ^*/hZ\bar{f}f$ couplings relevant to this process in the dimension-6 SMEFT. In particular, we propose a strategy to probe the anomalous $h Z_{\mu \nu}Z^{\mu \nu}$ and $h Z_{\mu \nu}\tilde{Z}^{\mu \nu}$ vertices at the percent level. We show that this can be achieved by resurrecting the interference term between the transverse $Zh$ amplitude, which receives contributions from the above couplings, and the dominant SM longitudinal amplitude. These contributions are hard to isolate without a knowledge of the analytical amplitude, as they vanish unless the process is studied differentially in three different angular variables  at the level of the $Z$-decay products. By also including the differential distributions   with respect to energy variables, we obtain projected bounds for the two other  tensor structures of the Higgs coupling to $Z$-bosons.
\end{abstract}

\maketitle

\section{Introduction}
The discovery of the Higgs boson \cite{Aad:2012tfa,Chatrchyan:2012xdj}, the first electroweak-scale scalar particle, marked the starting point for an ongoing extensive program to study its interactions with particles of the Standard Model to high precision~\cite{Khachatryan:2016vau, ATLAS-CONF-2018-031, Sirunyan:2018koj}. To perform this task, a theoretical framework was developed, compatible with high-scale UV completions of the Standard Model, which can mimic the kinematic impact of new resonances with masses beyond the energy-reach of the LHC, \textit{i.e.} the Standard Model Effective Field Theory (SMEFT) framework~\cite{Buchmuller:1985jz, Giudice:2007fh, Grzadkowski:2010es, Banerjee:2013apa, Elias-Miro:2013eta, Contino:2013kra, Gupta:2014rxa, Amar:2014fpa, Buschmann:2014sia, Craig:2014una, Ellis:2014dva, Ellis:2014jta, Banerjee:2015bla, Englert:2015hrx, Cohen:2016bsd, Ge:2016zro, Contino:2016jqw, Biekotter:2016ecg, deBlas:2016ojx, Denizli:2017pyu, Barklow:2017suo, Brivio:2017vri, Barklow:2017awn, Khanpour:2017cfq, Englert:2017aqb, Banerjee:2018bio, Biekotter:2018rhp, Goncalves:2018ptp,Freitas:2019hbk}. Different bases were proposed to parametrise the SMEFT operators, e.g. the SILH \cite{Giudice:2007fh} or Warsaw \cite{Grzadkowski:2010es} bases, each providing a generic and rather model-independent way to probe the couplings of the Standard Model.

An important class of interactions to probe the electroweak sector is the couplings of the Higgs boson to gauge bosons, and in particular to the $Z$-boson. There are 15 operators in the Warsaw basis at mass-dimension 6 that contribute to the $hZZ^*$ and $hZ\bar{f}f$ vertices (12 CP-even and 3 CP-odd operators). However, after electroweak symmetry breaking, these operators collectively only contribute to 4 interaction vertices for a given fermion, $f$. In the following section we explicitly show the relation between these dimension-6 operators and the  $hZZ^*/hZ\bar{f}f$ interaction vertices. 

Relying exclusively on the process $pp \to Z(\ell^+\ell^-)h(b\bar{b})$, we propose to exploit  differential distributions to constrain all 4 interaction vertices relevant to this process simultaneously. While there have been other studies devoted to this question~\cite{Godbole:2013lna, Godbole:2014cfa}, our approach is unique in that we systematically use our analytical knowledge of the squared amplitude to devise the experimental analysis strategy.  For the squared amplitude at the level of the $Z$-decay products, the three possible helicities of the intermediate $Z$-boson  give rise to 9 terms, each with a different angular  dependance. These 9 terms can be thought of as independent observables, each being sensitive to a different region of the final state's phase space. We assess which of these observables gets the dominant contribution from each of the 4 interaction vertices and thus devise a strategy to probe them simultaneously.  In particular, we isolate the interference term between the longitudinal and transverse amplitudes that  allows us to probe the  $h Z_{\mu \nu}Z^{\mu \nu}$ and $h Z_{\mu \nu}\tilde{Z}^{\mu \nu}$ vertices in a clean and precise way. 

This approach will be particularly useful for measurements during the upcoming high-luminosity runs of the LHC and at possible future high-energy colliders. It can be straightforwardly extended to other processes and different gauge bosons, and thus could play a crucial role in providing reliable and precise constraints in fits for effective operators. Exploiting and correlating different regions of phase space for individual processes can remove flat directions in the high-dimensional parameter space of effective theories. 

\section{Differential anatomy of $pp \to Z(\ell^+\ell^-)h(b\bar{b})$   in the SMEFT}
\label{sec:smeft}

Including all possible dimension 6 corrections, the most general $hZZ^*/hZ\bar{f}f$  vertex can be parameterised as follows (see for eg.~\cite{Isidori:2013cla, Gupta:2014rxa, Pomarol:2014dya})\footnote{Note that in the parametrisation of Ref.~\cite{Pomarol:2014dya, Banerjee:2018bio} both the custodial-preserving and breaking $hVV$ couplings contribute to $\delta \hat{g}^h_{ZZ}$.},
\begin{align}
\Delta {\cal L}^{hZ\bar{f}f}_6\supset   &\delta \hat{g}^h_{ZZ}\, \frac{2 m_Z^2}{v}h \frac{Z^\mu Z_\mu}{2}+ \sum_f g^h_{Zf}\,\frac{h}{v}Z_\mu \bar{f} \gamma^\mu f \nn\\&
+\kappa_{ZZ}\,\frac{h}{2v} Z^{\mu\nu}Z_{\mu\nu}+\tilde{\kappa}_{ZZ}\,\frac{h}{2v} Z^{\mu\nu}\tilde{Z}_{\mu\nu}.
\label{anam}
\end{align}
For a single fermion generation,  $f=u_L, d_L, u_R, d_R$ for corrections to the $pp \to Zh$ process and   $f=e_L,e_R$ for corrections to the $e^+e^- \to Zh$ process. The only model-independent bound on the above couplings is an ${\cal O}(10 \%)$ bound  from the global Higgs coupling fit~\cite{Khachatryan:2016vau, ATLAS-CONF-2018-031, Sirunyan:2018koj}. Translated to the above parametrisation, this would  constrain  a linear combination of the above couplings including, the leptonic $hZ\bar{f}f$ contact terms. If we limit ourselves to only universal corrections, we must replace the second term above by $h Z_\mu \partial_\nu Z^{\mu \nu}$, which can be written as a linear combination of the contact terms using the equations of motion. The above parametrisation is sufficient even if electroweak symmetry is non-linearly realised (see for eg.~\cite{Isidori:2013cga}). For the case of linearly realised electroweak symmetry, these vertices arise in the unitary gauge upon electroweak symmetry breaking. In the Warsaw basis~\cite{Grzadkowski:2010es}, we get the following contributions from the operators in  Table~\ref{opers}, 
 \bea
 \label{wilson}
 \delta \hat{g}^h_{ZZ}&=&\frac{v^2}{\Lambda^2}\left(c_{H\square}+\frac{3 c_{HD}}{4}\right)\nn\\
  g^h_{Zf}&=&- \frac{2 g}{\ctw}\frac{v^2}{\Lambda^2}(|T_3^f|c^{(1)}_{HF}-T_3^f c^{(3)}_{HF}+(1/2-|T_3^f|)c_{Hf})\nn\\
  \kappa_{ZZ}&=&\frac{2 v^2}{ \Lambda^2}(\ctw^2 c_{HW}+\stw^2 c_{HB}+ \stw \ctw c_{HWB})\nn\\
  \tilde{\kappa}_{ZZ}&=&\frac{2 v^2}{ \Lambda^2}(\ctw^2 c_{H\tilde{W}}+\stw^2 c_{H\tilde{B}}+ \stw \ctw c_{H\tilde{W}B}),
 \eea
where $F=Q (L)$ if $f$ is a quark (lepton).

 If electroweak symmetry is  linearly realised, there are additional constraints on the anomalous couplings in \eq{anam} because the same operators also contribute to different vertices already bounded by other  measurements. Using the formalism of BSM Primaries~\cite{Gupta:2014rxa}, we obtain, 
\bea
\delta \hat{g}^h_{ZZ}&=&\dfrac{2}{g^2}\left(\dfrac{\delta{g}^h_{VV}}{v}+\stw^2\delta g_1^Z-\ttw^2 \delta \kappa_\gamma \nn\right)\\
 g^h_{Zf}&=&\frac{2 g}{\ctw}Y_f \ttw^2 \delta \kappa_\gamma+2 \delta g^Z_{f}- \frac{2 g}{\ctw}(T^f_3 \ctw^2 + Y_f \stw^2)\delta g_1^Z \nn\\
{\kappa_{ZZ}} &=&\frac{{\delta \kappa_\gamma}}{2 c^2_{\theta_W}}+{ \kappa_{Z \gamma}} \frac{c_{2\theta_W}}{2 c^2_{\theta_W}}+{ \kappa_{\gamma \gamma}}\nn \\
{\tilde{\kappa}_{ZZ}} &=&\frac{{\delta \tilde{\kappa}_\gamma}}{2 c^2_{\theta_W}}+{ \tilde{\kappa}_{Z \gamma}} \frac{c_{2\theta_W}}{2 c^2_{\theta_W}}+{ \tilde{\kappa}_{\gamma \gamma}}
\label{linear}
\eea
The couplings in the right-hand side of the above equation are already constrained by LEP electroweak precision measurements or other Higgs measurements. The weakest constraint is on the triple gauge coupling $|\delta \kappa_\gamma|\lesssim 0.05$ \cite{LEP2}, which appears in the right-hand side of the first three equations above. This implies a 5 \% level bound on all the CP-even Higgs anomalous couplings. Note that it is extremely important to measure the anomalous couplings in the left-hand side of the above equations independently, despite these bounds. This is due to the fact that a verification of the above correlations can be used to  test whether electroweak symmetry is linearly or non-linearly realised. 
 
\begin{figure*}[!t]
\includegraphics[scale=0.4]{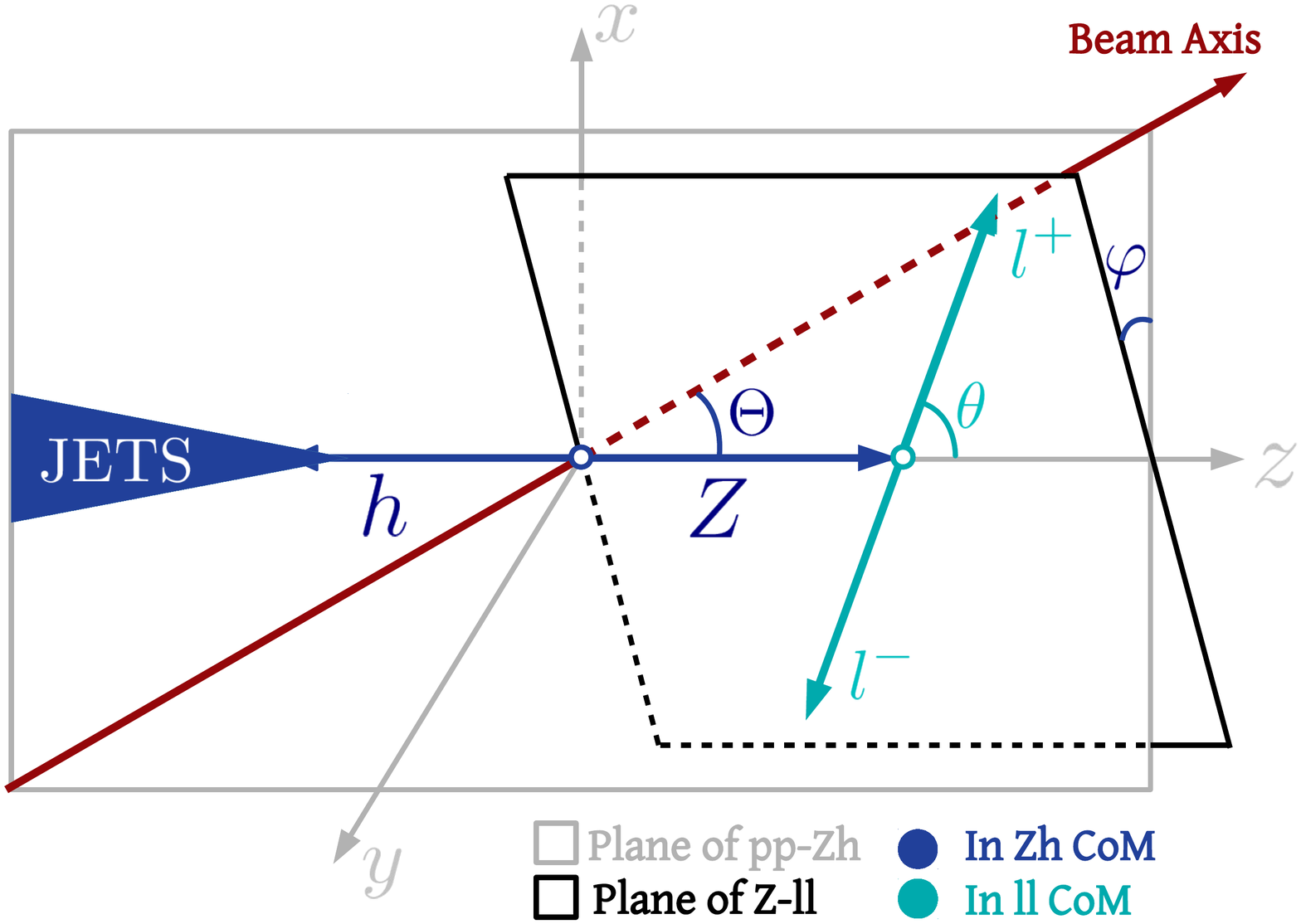}
\caption{Diagram showing the angles used to isolate the LT interference terms. Note that in fact two different frames of reference are represented: the CoM frame of the $Zh$ system (in which $\varphi$ and $\Theta$ are defined) and the CoM frame of the $Z$ (in which $\theta$ is defined). We define the Cartesian axes $\{x,y,z\}$ the $Zh$ centre-of-mass frame, with $z$ identified as the direction of the $Z$-boson; $y$ identified as the normal to the plane of the $Z$ and beam axis; finally $x$ is defined such that it completes the right-handed set. }
\label{fig:angles}
\end{figure*}

\begin{table}[t]
\small
\centering
\begin{tabular}{c}
\begin{tabular}{||c|c||}
\hline
&\\
                ${\cal O}_{H\square}=(H^\dagger H) \square (H^\dagger H)$&${\cal O}^{(3)}_{HL}=i H^\dagger \sigma^a \lra{D}_\mu H \bar{L}  \sigma^a \gamma^\mu L$ \\
\rule{0pt}{4ex} ${\cal O}_{HD}=(H^\dagger  {D}_\mu H)^*(H^\dagger  {D}_\mu H)$ &${\cal O}_{HB}= |H|^2 B_{\mu\nu}B^{\mu\nu}$\\
\rule{0pt}{4ex} ${\cal O}_{Hu}=i H^\dagger \lra{D}_\mu H \bar{u}_R  \gamma^\mu u_R$&${\cal O}_{HWB}=  H^\dagger \sigma^a H W^a_{\mu\nu}B^{\mu\nu}$\\
\rule{0pt}{4ex} ${\cal O}_{Hd}=i H^\dagger \lra{D}_\mu H \bar{d}_R  \gamma^\mu d_R$&${\cal O}_{H{W}}= |H|^2 W_{\mu\nu}{W}^{\mu\nu}$\\
\rule{0pt}{4ex} ${\cal O}_{He}=i H^\dagger \lra{D}_\mu H \bar{e}_R  \gamma^\mu e_R$&${\cal O}_{H\tilde{B}}= |H|^2 B_{\mu\nu}\tilde{B}^{\mu\nu}$ \\
\rule{0pt}{4ex} ${\cal O}^{(1)}_{HQ}=i H^\dagger  \lra{D}_\mu H \bar{Q}   \gamma^\mu Q$&${\cal O}_{H\tilde{W}B}=  H^\dagger \sigma^a H W^a_{\mu\nu}\tilde{B}^{\mu\nu}$ \\
\rule{0pt}{4ex} ${\cal O}^{(3)}_{HQ}=i H^\dagger \sigma^a \lra{D}_\mu H \bar{Q}  \sigma^a \gamma^\mu Q$&${\cal O}_{H\tilde{W}}= |H|^2 W^a_{\mu\nu}\tilde{W}^{a \mu\nu}$ \\
\rule{0pt}{4ex} ${\cal O}^{(1)}_{HL}=i H^\dagger  \lra{D}_\mu H \bar{L}   \gamma^\mu L$& \\
&\\
\hline
 \end{tabular}
\end{tabular}
\caption{Dimension-6 operators in the Warsaw basis that contribute to the anomalous $hZZ^*/hZ\bar{f}f$ couplings in \eq{anam}. Here $Q$ and $L$ are the quark and lepton doublets. For other details regarding the notation see~\cite{Grzadkowski:2010es}.}
\label{opers}
\end{table}
 The main objective of this work is to study the Higgs-strahlung process differentially with respect to energy and angular variables in order to individually constrain all the above anomalous couplings. To isolate the effects of the different couplings above it is most convenient to use the helicity amplitude formalism. At the 2$\to$2 level,  $f(\sigma)\bar{f}(-\sigma) \to Zh$, these helicity amplitudes are given by,

 \begin{widetext}
\bea
\label{amp}
\hspace{-5mm}&&{\cal M}_\sigma^{\lambda=\pm}= \sigma \frac{1+ \sigma \lambda \cos \Theta}{\sqrt{2}}\frac{g g^Z_f}{\ctw}\frac{m_Z}{\sqrt{\hat{s}}}\Bigg[1+\left(\frac{g^h_{Zf}}{g^Z_f}+  \kappa_{ZZ}- i \lambda \tilde{\kappa}_{ZZ}\right)  \frac{\hat{s}}{2 m_Z^2}  \Bigg]\nn\\
&&{\cal M}_\sigma^{\lambda=0}= -\sin \Theta\frac{g g^Z_f}{2\ctw}\Bigg[1+\delta \hat{g}^h_{ZZ}+  2\kappa_{ZZ}+\frac{{g}^h_{Zf}}{g^Z_{f}} \left(-\frac{1}{2}+\frac{\hat{s}}{2 m_Z^2}\right) \Bigg],
\eea
\end{widetext}
 where  $\lambda=\pm 1$ and $\sigma=\pm 1$ are, respectively, the helicities of the $Z$-boson and initial-state fermions, and $g^Z_{f}=g(T_3^f-Q_f \stw^2)/\ctw$; $\sqrt{\hat{s}}$ 
 is the partonic centre-of-mass energy.  We have kept only terms with the highest powers of $\gamma={\sqrt{\hat{s}}}/(2 m_Z)$  in the expressions above, both for the SM and EFT contributions. The neglected terms are smaller at least by a factor of $4 m_Z^2/\hat{s}$.  An exception is the next-to-leading EFT contribution for the $\lambda=0$ mode, which we retain in order to keep the leading effect amongst  the terms proportional to $\delta \hat{g}^h_{ZZ}$ term. For the full expressions see~\cite{Nakamura:2017ihk}. The above expressions assume that the quark  moves in the positive $z$ direction and the opposite case, where the antiquark direction coincides with the positive $z$ direction, can be obtained by replacing $\sigma\to- \sigma$. Here and in what follows, unless explicitly mentioned,  our analytical expressions hold for both quark and leptonic initial states.

At high energies the dominant EFT correction   is to the longitudinal mode ($\lambda=0$). For the $pp\to Zh$ process at the LHC, a linear combination of the four contact-term couplings, $g^h_{Zf}$, enters the EFT correction to the longitudinal cross-section. This linear combination, given by, 
 \beq
g^h_{Z\textbf{p}}=g^h_{Zu_L} -0.76~g^h_{Zd_L}   - 0.45~g^h_{Zu_R} + 0.14~g^h_{Zd_R}  \,,
\label{compdir}
\eeq 
arises from the inability to disentangle the polarisation of the initial partons, and that the luminosity ratio of up and down quarks remains roughly constant over the relevant energy range~\cite{Banerjee:2018bio}. As shown in~\cite{Banerjee:2018bio}, by constraining  these  deviations that grow with energy, one can obtain  strong  per-mille-level bounds on $g^h_{Z\textbf{p}}$, even with 300 fb$^{-1}$ LHC data. The corrections to the longitudinal mode are also related to longitudinal double gauge-boson production due to the Goldstone boson equivalence theorem~\cite{Franceschini:2017xkh}.
 
The unique signatures of the $\kappa_{ZZ},\tilde{\kappa}_{ZZ}$ couplings arise from their contributions to the transverse $Zh$ mode ($\lambda=\pm 1$), which  in the SM  is subdominant at high energies. The corrections to the transverse mode are hard to probe as this mode does not interfere with the dominant SM longitudinal mode. However, the longitudinal-transverse (LT) interference term is present at the level of the $Z$-decay products and vanishes only if we integrate  inclusively over their phase space.\footnote{This is analogous to the case of double gauge-boson production where a similar situation arises for certain triple gauge-boson deformations that contribute   to helicity amplitudes that are  subdominant in the SM~\cite{Hagiwara:1986vm, azatov, panico, azatov2}.} To recover this interference term and, in general, to maximally discriminate the transverse mode from the longitudinal mode, we must utilise the full dependence of the differential cross-section on $\Theta$, and the angular variables related to the $Z$ decay products (as defined in Fig.~\ref{fig:angles}). Analytically, the amplitude can be most conveniently written in terms of $\hat{\varphi}$, the azimuthal angle of the positive-helicity lepton and $\hat{\theta}$, its polar angle in the $Z$ rest frame. In terms of these variables the  amplitude is given by,
\bea
\label{helamp}
{\cal A}_h(\hat{s},  \Theta, \hat{\theta}, \hat{\varphi})=\frac{-i\sqrt{2} g^Z_\ell}{\Gamma_Z}\sum_\lambda {\cal M}_\sigma^\lambda (\hat{s},\Theta)d^{J=1}_{\lambda,1}(\hat{\theta}) e^{i \lambda \hat{\varphi}}
\eea
where  $d^{J=1}_{\lambda,1}(\hat{\theta})$ are the Wigner functions (see for eg.~\cite{Panico:2017frx}), $\Gamma_Z$ is the $Z$-width and $g^Z_{\ell}=g(T_3^\ell-Q_\ell\stw^2)/\ctw$. Given that the polarisation of the final state lepton is not experimentally accessible, we express the squared amplitude (after summing over the final lepton polarisations) in terms of ${\theta}$ and $\varphi$, the analogous angles for the positively-charged lepton, 
\bea
\label{sumchiral}
\sum_{L,R}|{\cal A}(\hat{s}, \Theta,{\theta}, {\varphi})|^2&=&\alpha_L |{\cal A}_h (\hat{s},\Theta,{\theta}, {\varphi})|^2\nn\\&+&\alpha_R|{\cal A}_h (\hat{s},\Theta,\pi-{\theta}, \pi+{\varphi})|^2,
\eea
where $\alpha_{L,R}= (g^Z_{l_{L,R}})^2/[(g^Z_{l_L})^2+(g^Z_{l_R})^2]$ is the fraction of  $Z\to \ell^+ \ell^-$ decays to leptons with left-handed (right-handed) chiralities. The above equation follows from the fact that for left-handed chiralities, the positive-helicity lepton is  the positively-charged lepton, whereas it is the negatively-charged lepton for right-handed chiralities, so that for the latter case $(\hat{\theta}, \hat{\varphi})=(\pi-{\theta}, \pi+{\varphi})$. Using equations (\ref{amp},\ref{helamp},\ref{sumchiral}) one can write the full angular dependance of the squared amplitude, giving nine  angular  functions of $ \Theta,{\theta}$ and ${\varphi}$ (see also~\cite{Collins:1977iv, Hagiwara:1984hi, Goncalves:2018ptp}),

\begin{table}[t]
\small
\centering
\begin{tabular}{||c|c||}
\hline
                $a_{LL}$           & $\frac{{\cal G}^2}{4}\Big[1+2\delta \hat{g}^h_{ZZ}+  4\kappa_{ZZ}+\frac{{g}^h_{Zf}}{g^Z_{f}}(-1+4 \gamma^2)\Big]$\\
\rule{0pt}{4ex} $a^1_{TT}$         & $\frac{{\cal G}^2 \sigma \epsilon_{LR}}{2\gamma^2}\Big[1+4\left(\frac{{g}^h_{Zf}}{g^Z_{f}}+\kappa_{ZZ}\right)\gamma^2\Big]$\\
\rule{0pt}{4ex} $a^2_{TT}$         & $\frac{{\cal G}^2}{8\gamma^2}\Big[1+4\left(\frac{{g}^h_{Zf}}{g^Z_{f}}+\kappa_{ZZ}\right)\gamma^2\Big]$\\
\rule{0pt}{4ex} $a^1_{LT}$         & $-\frac{{\cal G}^2 \sigma\epsilon_{LR}}{2\gamma}\Big[1+2\Big(\frac{2{g}^h_{Zf}}{g^Z_{f}}+\kappa_{ZZ}\Big){\gamma^2}\Big]$\\
\rule{0pt}{4ex} $a^2_{LT}$         & $-\frac{{\cal G}^2}{2\gamma}\Big[1+2\Big(\frac{2{g}^h_{Zf}}{g^Z_{f}}+\kappa_{ZZ}\Big){\gamma^2}\Big]$\\
\rule{0pt}{4ex} $\tilde{a}^1_{LT}$ & $-{\cal G}^2 \sigma\epsilon_{LR} \tilde{\kappa}_{ZZ} \gamma$\\
\rule{0pt}{4ex} $\tilde{a}^2_{LT}$ & $-{\cal G}^2 \tilde{\kappa}_{ZZ}\gamma$\\
\rule{0pt}{4ex} $a_{TT'}$          &$\frac{{\cal G}^2}{8\gamma^2}\Big[1+4\left(\frac{{g}^h_{Zf}}{g^Z_{f}}+\kappa_{ZZ}\right)\gamma^2\Big]$\\
\rule{0pt}{4ex} $\tilde{a}_{TT'}$  & $\frac{{\cal G}^2}{2} \tilde{\kappa}_{ZZ}$\\
\hline
 \end{tabular}
\caption{Contribution of the different anomalous couplings in \eq{anam} to the angular coefficients in \eq{nine} up to linear order. The above expressions hold for the case that the initial quark direction coincides with the positive $z$ direction. To obtain the final expressions relevant for the LHC we must average over this and the other other possibility  that  the antiquark moves in the positive $z$ direction (which is obtained by replacing  $\sigma \to -\sigma$ ). This leads to a vanishing   $a^1_{TT},a^1_{LT}$ and $\tilde{a}^1_{LT}$ while keeping the other coefficients unchanged. Contributions subdominant in $\gamma^{-1}={2 m_Z}/\sqrt{s}$ have been neglected, with the exception of the next-to-leading EFT contribution to $a_{LL}$, which we retain in order to keep the leading effect of the $\delta \hat{g}^h_{ZZ}$ term. The terms neglected are smaller by at least  a factor of $1/\gamma^2$. Here $\epsilon_{LR}= \alpha_L-\alpha_R,~{\cal G}= g g^Z_f \sqrt{(g^Z_{l_L})^2+(g^Z_{l_R})^2}/(\ctw \Gamma_Z)$ and $ \Gamma_Z$ is the $Z$-width. The SM part our results are in complete agreement with~\cite{barger}. }
\label{coefs}
\end{table}

\begin{align}
\label{nine}
&\sum_{L,R}|{\cal A}(\hat{s}, \Theta,{\theta}, {\varphi})|^2 = a_{LL} \sin^2 \Theta \sin^2 \theta +a^1_{TT}\cos \Theta \cos \theta
\nn\\&+a^2_{TT}(1+ \cos^2 \Theta)(1+ \cos^2 \theta)+\cos \varphi  \sin \Theta\sin \theta\nn\\&\times( a^1_{LT}+a^2_{LT} \cos \theta \cos\Theta)+ \sin \varphi\sin \Theta \sin \theta\nn\\&\times( \tilde{a}^1_{LT}+ \tilde{a}^2_{LT} \cos \theta \cos\Theta)+  a_{TT'}\cos 2\varphi  \sin^2 \Theta \sin^2 \theta\nn\\&+ \tilde{a}_{TT'}\sin 2\varphi  \sin^2 \Theta \sin^2 \theta.
\end{align}
The subscripts of the above coefficients denote the $Z$-polarisation of the two interfering amplitudes, with $TT'$ denoting the interference of two transverse amplitudes with opposite polarisations. These  coefficients should be thought of as independently-measurable observables.

Expressions for the nine coefficients above in terms of the anomalous couplings are given in  Table~\ref{coefs}.  The expression in Table~\ref{coefs}  utilise \eq{amp} which assumes that the initial quark direction coincides with the positive $z$ direction. To obtain the final expressions relevant for the LHC we must average over this and the other other possibility  that  the antiquark moves in the positive $z$ direction (which is obtained by replacing  $\sigma \to -\sigma$ in \eq{amp} and Table~\ref{coefs}). This leads to a vanishing   $a^1_{TT},a^1_{LT}$ and $\tilde{a}^1_{LT}$ while keeping the other coefficients unchanged. Notice that powers of $\gamma=\sqrt{\hat{s}}/(2 m_Z)$ lead to a parametric enhancement in some of the contributions to the coefficients. The  dominant EFT contribution is that of $g^h_{Zf}$ to  $a_{LL}$. This coefficient also receives a subdominant contribution from $\delta \hat{g}^h_{ZZ}$.  A linear combination of  $\kappa_{ZZ}$ and $g^h_{Zf}$  gives the dominant contribution   to 3 of the remaining coefficients, namely: $a^2_{TT}, a^2_{LT}$ and $a^1_{TT'}$. Similarly, $\tilde{\kappa}_{ZZ}$ is the only coupling that contributes to the 2 non-zero  CP-violating parameters: $ \tilde{a}^2_{LT}$ and $\tilde{a}_{TT'}$. 

As anticipated, the parametrically-largest contribution is to the LT  interference terms, 
 \beq
 \label{ltint}
  \frac{a^2_{LT}}{4}\cos \varphi \sin 2\theta  \sin 2\Theta+ \frac{\tilde{a}^2_{LT}}{4}\sin \varphi \sin 2\theta \sin 2\Theta.
 \eeq
 By looking at the dependance of  $a_{LL},a^2_{LT}$ and $\tilde{a}^2_{LT}$ on the initial quark helicity, $\sigma$, we see that the
  linear combination of $g^h_{Zf}$ couplings that enters $a^2_{LT}$ and $\tilde{a}^2_{LT}$ for the $pp \to Zh$ process is again $g^h_{Z\textbf{p}}$ defined in \eq{compdir}. Once $g^h_{Z\textbf{p}}$ is very-precisely constrained by constraining $a_{LL}$ at high energies, one can separate the contribution of $\kappa_{ZZ}$ to the 2 coefficients mentioned above. In the following sections we isolate these terms in our experimental analysis in order to constrain    ${\kappa}_{ZZ}$ and $\tilde{\kappa}_{ZZ}$.  Notice  that the above terms give no contribution if we integrate inclusively over either $\Theta, \theta$ or $\varphi$. It is therefore highly non-trivial to access the LT interference term if one is not guided by the analytical form above.

Finally, we constrain $\delta\hat{g}^h_{ZZ}$. This coupling only rescales  the SM $hZZ$ coupling and  hence all SM differential distributions. In order to constrain  this coupling one needs to access its contribution to $a_{LL}$, which is subdominant  in $\gamma$  (see Table~\ref{coefs}).    Ideally, one can perform a fit to the differential distribution with respect to $\hat{s}$ to extract both the dominant and subdominant pieces.   In this work we will study the differential distribution  with respect to $\hat{s}$ in two ranges, a low and high  energy range, in order  to individually constrain  both $g^h_{ZZ}$ and $g^h_{Z\textbf{p}}$ (see Sec~\ref{sec:analysis}).

We have thus identified four observables to constrain the four anomalous couplings in \eq{anam}; these are: the differential $pp \to Zh$ cross-section with respect to $\hat{s}$ at high  and low energies, and the angular observables $a^2_{LT}$ and $\tilde{a}^2_{LT}$. While we have chosen the observables that receive the largest EFT corrections parametrically, ideally one should use all the information contained in the nine coefficients in \eq{nine} (especially in the unsuppressed $a^2_{TT}, a_{TT'}$ and $\tilde{a}_{TT'}$) to obtain the strongest possible constraints on the Higgs anomalous couplings in \eq{anam}. We leave this for future work.

 We have so far considered only the effect of the anomalous Higgs couplings in \eq{anam}. The $pp \to Z(\ell^+\ell^-) h(b\bar{b})$ process, however, also gets contributions~\cite{Gupta:2014rxa} from operators that rescale  the $hb\bar{b}$ and $Z\bar{f}f$ couplings (that we parametrise here by $\delta \hat{g}^h_{b\bar{b}}$ and $\delta \hat{g}^Z_{f}$ respectively) and  from the vertices, 
\bea
\kappa_{Z\gamma}\,\frac{h}{v} A^{\mu\nu}Z_{\mu\nu}+\tilde{\kappa}_{Z\gamma}\,\frac{h}{v} A^{\mu\nu}\tilde{Z}_{\mu\nu}.
\eea
The effect of these couplings can be incorporated by simply replacing in all our expressions, 
\bea
\delta\hat{g}^h_{ZZ} &\to& \delta\hat{g}^h_{ZZ}+\delta \hat{g}^h_{b\bar{b}} + \delta \hat{g}^Z_{f}, \nn\\
\kappa_{ZZ} &\to& \kappa_{ZZ}+\frac{Q_f e}{g^Z_f}\kappa_{Z\gamma},\nn\\
\tilde{\kappa}_{ZZ} &\to& \tilde{\kappa}_{ZZ}+\frac{Q_f e}{g^Z_f}\tilde{\kappa}_{Z\gamma},
\eea
where for the last two replacements we have assumed  $\hat{s}\gg m_Z^2$. At the $pp \to Zh$ level, the last two replacements become $\kappa_{ZZ} \to \kappa_{ZZ}+0.3~\kappa_{Z\gamma}$, $\tilde{\kappa}_{ZZ} \to \tilde{\kappa}_{ZZ}+0.3~\tilde{\kappa}_{Z\gamma}$. These degeneracies can be resolved straightforwardly by including LEP $Z$-pole data and information from other Higgs-production and decay channels.

\section{Analysis and Results}
\label{sec:analysis}

The following analysis is performed for $\sqrt{s} = 14$ TeV. We base our analysis strategy on the one described in~\cite{Banerjee:2018bio}.  Our signal comprises of $Zh \to \ell^+ \ell^- b \bar{b}$ production from a pair of quarks and gluons with the former being the dominant contribution. We consider the dominant backgrounds, which consist of the SM $Zh$ production decaying in the same final state, $Zb\bar{b}$ (where the subdominant gluon-initiated case is also taken into account) and $Z+$jets (where jets include $c$ quarks as well, but are not explicitly tagged), where the light jets can fake as $b$-tagged jets. We also consider the leptonic mode of the $t\bar{t}$ process. 

In order to isolate events where a  a boosted Higgs boson gives rise to the $b \bar{b}$  pair,  from significantly larger QCD backgrounds, we resort to a fat jet analysis instead of a resolved analysis. For the fat jet analysis, we follow the BDRS technique~\cite{Butterworth:2008iy, Soper:2010xk, Soper:2011cr} with small alterations in order to maximise the sensitivity. The details of the analysis are presented in Appendix~\ref{app:1}. Using a multivariate analysis (MVA), described in detail in Appendix~\ref{app:1}, we enhance  the ratio  of   SM $Zh(b\bar{b})$ to $Zb\bar{b}$   events from a factor of  0.02 to  about 0.2, still keeping around 500   $Zh(b\bar{b})$ events with 3 ab$^{-1}$ data for a certain value of the MVA score. For a tighter MVA cut, we increase this ratio even further (to about  $\sim 0.5$). We will use booth these cuts in what follows.

\begin{figure}[!t]
\includegraphics[scale=0.55]{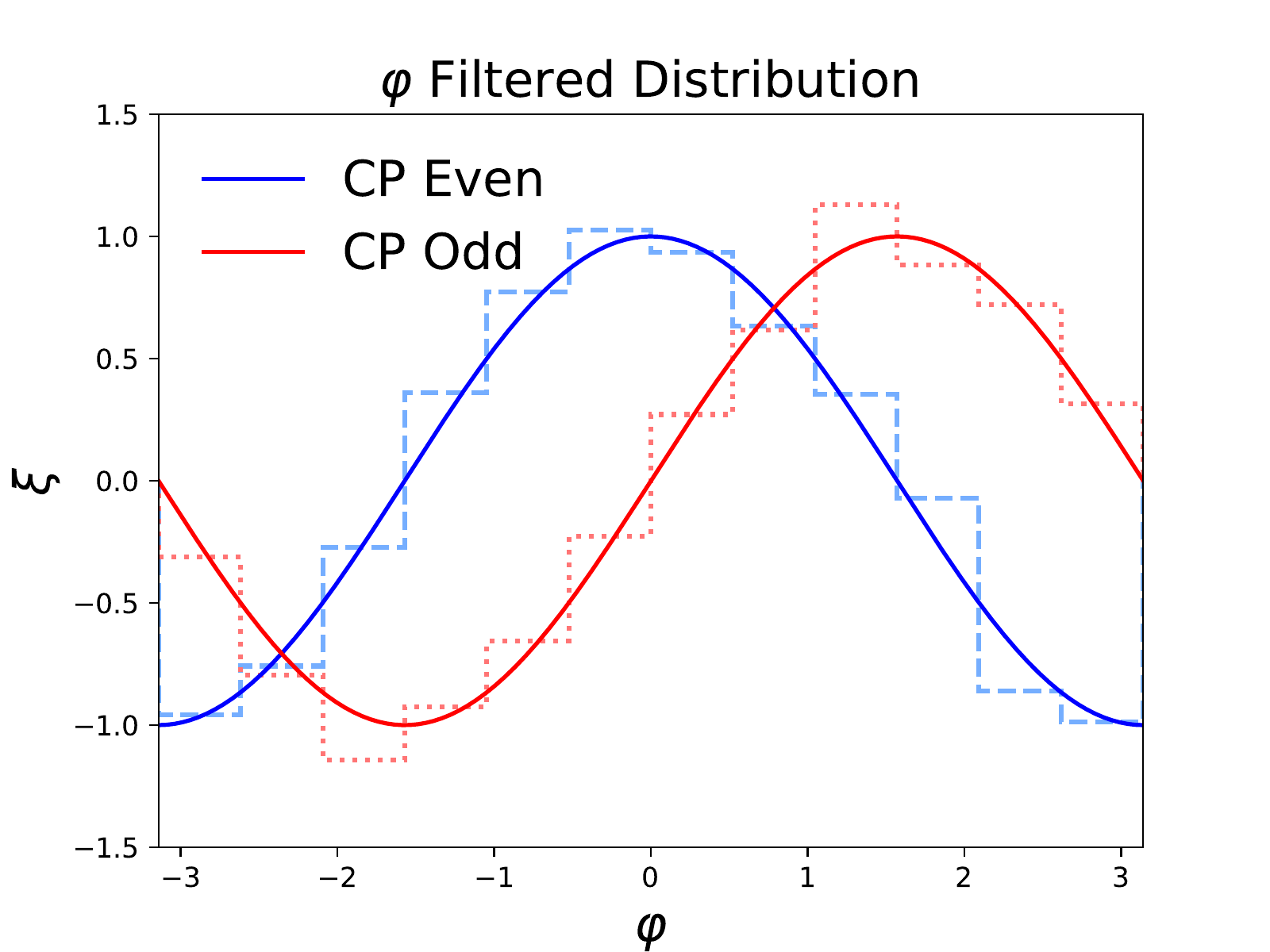}
\caption{We show here the  LT interference terms in \eq{ltint}  for  CP odd and even (from the couplings $\kappa_{ZZ}$  and $\tilde{\kappa}_{ZZ}$) EFT contributions. To visualise these contributions we have carried out a weighted integration giving  a negative weight whenever  the product $\sin 2 \Theta \sin 2\theta$ is negative. This yields the asymmetry variable, $\xi(\varphi)$, with the expected  $\cos \varphi$ ($\sin \varphi$)  dependance for  the $\kappa_{ZZ}$ ($\tilde{\kappa}_{ZZ}$) contribution shown above. }
\label{sincos}
\end{figure}

We now use the four differential  observables identified in Sec.~\ref{sec:smeft} to obtain sensitivity projections for the four anomalous couplings in \eq{anam}. We will determine the value of a given anomalous coupling that can be excluded at   68\% CL level, assuming that the observed number of events agrees with the SM.  For a given value of the anomalous couplings, one can estimate the cutoff for our EFT by putting the Wilson Coefficients, $c_i=1$, in \eq{wilson}. We will ignore in our analysis any event with a $Zh$ invariant mass, $M_{Zh}$, larger than the estimated cutoff.

\paragraph{High energy   $M_{Zh}$ distribution:} As already discussed, by just looking at the  tail of the  distribution with respect to $M_{Zh}$, one can constrain the leading energy enhanced contribution to $a_{LL}$ induced by  $g^h_{Z \textbf{p}}$. The analysis in Ref.~\cite{Banerjee:2018bio} reveals that one can obtain the following per-mille-level bound with 3 ab$^{-1}$ data,
\beq
\label{strong}
|g^h_{Z \textbf{p}}| <5 \times 10^{-4}.
\eeq

\paragraph{Low energy $M_{Zh}$ distribution:} 

Once the $M_{Zh}$ distribution at high energies has been used to obtain the strong bound on $g^h_{Z \textbf{p}}$ in \eq{strong}, one can use the lower energy bins to constrain the subdominant contribution of $\delta \hat{g}^h_{ZZ}$ (see Table~\ref{coefs}). We have checked, for instance, that for  $M_{Zh}< 950$ GeV, values of $g^h_{Z \textbf{p}}$   smaller than the bound in \eq{strong} have a negligible contribution. Using the sample with the tighter MVA cut,  we distribute the data into 100 GeV $M_{Zh}$ bins. We then construct a bin-by-bin $\chi^2$ function, where for each bin we add in quadrature a 5 \% systematic error to the statistical error. Energy-independent corrections from $\kappa_{ZZ}$ to $a^2_{TT}$ (see Table~\ref{coefs}) are also of the same order as the $\delta \hat{g}^h_{ZZ}$ contribution. Including these corrections,  for an integrated luminosity of 3 ab$^{-1}$, we finally obtain a bound on the linear combination,
 \beq
 \label{ghzz}
 -0.06<  \delta\hat{g}^h_{ZZ} +3.5~\kappa_{ZZ}  <0.07,
 \eeq
  where we have ignored any events with $M_{Zh}>950$ GeV, the EFT cutoff estimated as discussed above. The precise linear combination that appears above is of course dependent on the choice of our cuts and has been obtained numerically after our collider analysis.

\paragraph{The LT interference terms $a^2_{LT}$ and $\tilde{a}^2_{LT}$:} We want to isolate the terms in \eq{ltint}, which vanish upon an inclusive integration over either $\Theta, \theta$ and $\varphi$. To visually show the impact of turning on the couplings $\kappa_{ZZ}$ and $\tilde{\kappa}_{ZZ}$,  we carry out a weighted integration, which gives an event a weight equal to the sign of $\sin 2 \Theta \sin 2\theta$. This yields an asymmetry variable, $\xi(\varphi)$, which is expected to have a $\cos \varphi$ ($\sin \varphi$)  dependance for  the $\kappa_{ZZ}$ ($\tilde{\kappa}_{ZZ}$) contribution. We show a normalised histogram for $\xi$ with respect to $\varphi$ in Fig.~\ref{sincos}.  As expected from Table~\ref{coefs}, we also find an SM contribution to $a^2_{LT}$ with respect to which  the EFT contribution grows as  $2 \kappa_{ZZ}\gamma^2$ at high energies. The contribution of the remaining background to $a^2_{LT}$ is about four times the SM contribution. 

 For the final extraction of $a^2_{LT}$ and $\tilde{a}^2_{LT}$ we convolute the observed angular distribution in each energy bin with the weight functions $\cos \varphi \sin 2\theta  \sin 2\Theta$ and $\sin \varphi \sin 2\theta  \sin 2\Theta$ respectively. It can be checked  that this uniquely isolates $a^2_{LT}$ and $\tilde{a}^2_{LT}$ respectively amongst  the nine coefficients of \eq{nine}. Using the fact that these coefficients depend linearly on $\kappa_{ZZ}$ and $\tilde{\kappa}_{ZZ}$ (assuming again that $g^h_{Z \textbf{p}}$ is precisely constrained), respectively, we translate the values of the coefficients to these anomalous couplings. In practice to carry out the above convolution we perform a weighted sum over the simulated Monte Carlo events with the above weights where we use the sample with the looser MVA cut. To estimate the uncertainties we split out Monte Carlo sample into multiple smaller samples each with the expected number of events at 3 ab$^{-1}$ and find the value of $\kappa_{ZZ}$ and $\tilde{\kappa}_{ZZ}$ in each case. We finally obtain the 1$\sigma$ bound:
 \bea
 \label{result}
-0.07<  &\kappa_{ZZ}&  <0.07\nn\\
-0.07<  &\tilde{\kappa}_{ZZ}&  <0.07.
\label{finalb}
 \eea
 Again we ignore events with $M_{Zh}$ larger than the cutoff estimated by the procedure discussed above. In any case our result is not too dependent on  this procedure as we obtain  maximal sensitivity  from events in the $450\lesssim M_{Zh}\lesssim850$ GeV range, which is safely below the estimated cut-off.

We now compare our final bounds in \eq{finalb} with other existing projections on the measurement of $\kappa_{ZZ}$ and $\tilde{\kappa}_{ZZ}$. The projections of Ref.~\cite{caola}  from the $h \to ZZ \to 4l$ process at 3 ab$^{-1}$ using the matrix element method are  $|\kappa_{ZZ}|  <0.04$ and  $|\tilde{\kappa}_{ZZ}|  <0.09$.\footnote{The projections in Refs.~\cite{Godbole:2014cfa,caola} from $pp \to Zh$ are unfortunately not comparable to ours as these studies include high energy regions of the phase space  where the EFT rate is many times that of the SM. These results are thus not compatible with our assumption of ${\cal O}(1)$ Wilson coefficients.} Bounds on $\kappa_{ZZ}$ can also be obtained using \eq{linear} and the 3 ab$^{-1}$ projection from diboson production  $\delta \kappa_\gamma \lesssim 0.01$~\cite{montull}. While this results in the more stringent bound  $\kappa_{ZZ} \lesssim 0.01$, it assumes that electroweak symmetry is linearly realised. If, instead, we want to establish (or disprove) that electroweak symmetry is linearly realised using precision Higgs physics, it is essential to measure all the couplings in \eq{linear} independently.

%
%

\section{Conclusion}

     As we enter the era of higher energies and luminosities,  time has come to shift from using only rate information to  performing  differential studies that utilise more sophisticated kinematical observables. In this work we have shown how a differential study of the $pp \to Z(\ell^+\ell^-)h(b\bar{b})$ process can completely resolve the tensor structure of the  $hZZ^*/hZ\bar{f}f$ contributions in the dimension 6 SMEFT (see \eq{anam} and \eq{wilson}).
     
      To achieve this, we have studied analytically the full differential cross section in the SMEFT (see \eq{nine}).  This has enabled us to identify differential observables that get leading contributions from  the different anomalous vertices. Of the four possible anomalous Higgs couplings relevant to this process,  $g^h_{Z\textbf{p}}$ and $\delta\hat{g}^h_{ZZ}$ can be constrained   using the differential distribution with respect to the $Zh$ invariant mass. The  leading contributions from  $\kappa_{ZZ}$ and $\tilde{\kappa}_{ZZ}$ are much more elusive. This is because the above couplings give corrections only to transverse $Zh$ production, which does not interfere with the dominant  SM amplitude for the longitudinal mode. The interference term (see \eq{ltint})  can be recovered at the level of the $Z$ decay products but only if we perform the analysis differentially in three angular variables. We ultimately show that at the high luminosity LHC one can constrain $g^h_{Z\textbf{p}}$ at the per-mille-level, $\delta\hat{g}^h_{ZZ}$ at the 5 \% level  and the couplings $\kappa_{ZZ}$ and $\tilde{\kappa}_{ZZ}$ at the percent level (see \eq{strong}, \eq{ghzz} and \eq{result}, respectively).

In this study   we have identified  4 optimal observables in order to  obtain simultaneous bounds on  the 4  anomalous Higgs couplings. Our sensitivity estimates are thus conservative, as there are many more observables we have not considered. Even for the observables we consider, our analysis does not utilise the full angular shape information. There is thus the possibility that significantly stronger bounds can be obtained if the   full differential  information contained in the  matrix element squared (see Table~\ref{coefs}) is extracted  by using, for example, the method of angular  moments (see eg. Refs.~\cite{Dighe:1998vk,James, Beaujean:2015xea}) or  advanced machine-learning tools.  The approach advocated here is equally applicable to future leptonic colliders where it can be of even greater importance as, in this case,   Higgs-strahlung is among the dominant Higgs production modes.
  
\subsection*{Acknowledgments}
We thank Amol Dighe and Shilpi Jain for helpful discussions. S.B. is supported by a Durham Junior Research Fellowship COFUNDed by Durham University and the European Union, under grant agreement number 609412.
\appendix

\section{Details of the collider analysis}
\label{app:1}

Our analysis setup can be described in the following steps. We create our model containing all the effective vertices using \texttt{FeynRules}~\cite{Alloul:2013bka} and obtain the \texttt{UFO}~\cite{Degrande:2011ua} model implementation which is then fed into the \texttt{MG5$\_$aMC@NLO}~\cite{Alwall:2014hca} package used to generate all the signal and background samples at leading order (LO). For the loop-induced processes, we perform the decays using \texttt{MadSpin}~\cite{Frixione:2007zp,Artoisenet:2012st}. Next we hadronise and shower the events using the \texttt{Pythia 8}~\cite{Sjostrand:2001yu,Sjostrand:2014zea} framework. Finally, we perform a simplified detector simulation, which we discuss shortly.

Since we are looking into a boosted topology, we generate the $Zh$ and $Zb\bar{b}$ samples with the following generation level cuts: 
$p_{T,(j,b)} > 15$ GeV, $p_{T,\ell} > 5$ GeV, $|y_j| < 4$, $|y_{b/\ell}| < 3$, $\Delta R_{b\bar{b}/bj/bl} > 0.2$, $\Delta R_{\ell^+\ell^-} > 0.15$, $70 \; \text{GeV} < m_{\ell \ell} < 110$ GeV, $75 \; \text{GeV} < m_{b\bar{b}} < 155$ GeV and $p_{T,\ell^+\ell^-} > 150$ GeV. Moreover, these processes are generated with an additional parton upon using the matrix element (ME) parton shower (PS) merging in the MLM merging scheme~\cite{Mangano:2006rw}. The events in the $Z+$ jets channel are generated without the cut on the invariant mass of the jets and upon merging with up to three ME partons. All our event generations are at LO. Hence, in order to taken into account higher-order QCD corrections, we use next-to-leading order (NLO) $K$-factors. For the $qq$ initiated $Zh$ samples, we include a bin-by-bin NLO corrected $K$-factor in the reconstructed $M_{Zh}$ (invariant mass of the double $b$-tagged filtered fat jet and the two isolated leptons) distribution for both the SM and the EFT signal~\cite{Greljo:2017spw}. For the $gg$ initiated counterpart, we multiply the LO cross-section by a flat $K$-factor of 2~\cite{Altenkamp:2012sx}. For the tree-level $Zb\bar{b}$ and $Z+$ jets backgrounds, we respectively use $K$-factors of 1.4 (computed in the \texttt{MG5$\_$aMC@NLO} framework) and 0.91~\cite{Campbell:2002tg}. Finally, we consider an NLO correction of 1.8~\cite{Alioli:2016xab} for the $gg$ initiated $Zb\bar{b}$ process. Further electroweak backgrounds \cite{Campanario:2010hp} are found to be small.

As mentioned in above, we use the BDRS technique to optimise our signal yield. The BDRS technique reconstructs jets upon using the Cambridge-Aachen (CA) algorithm~\cite{Dokshitzer:1997in,Wobisch:1998wt} with a large cone radius in order to contain all the decay products ensuing from the relevant resonance. One then looks at the substructure of this fat jet by working backwards through the jet clustering. The algorithm requires us to stop when a substantial \textit{mass drop}, $m_{j_1} < \mu m_j$ with $\mu=0.66$, (where $m_j$ is the mass of the fatjet) occurs for a reasonably symmetric splitting, $$\frac{\text{min}(p_{T,j_1}^2,p_{T,j_2}^2)}{m_j^2}\Delta R_{j_1,j_2}^2>y_{\text{cut}},$$ with $y_{\text{cut}} = 0.09$. If the aforementioned criteria is not met, one removes the softer subjet, $j_2$ and $j_1$ is subjected to the above criteria. This iterative algorithm stops once one finally obtains two subjets, $j_1$ and $j_2$ which satisfy the mass drop criteria. In order to improve the reconstruction, the mass drop criteria is combined with the \textit{filtering} algorithm. For this step, the two subjets $j_1$ and $j_2$ are further combined using the CA algorithm upon using a cone radius of $R_{\text{filt}} = \text{min}(0.3, R_{b\bar{b}}/2)$. Finally, only the hardest three filtered subjets are considered to reconstruct the resonance. However, in our study we find that using $R_{\text{filt}} = \text{max}(0.2, R_{b\bar{b}}/2)$ acts as a better choice in reducing backgrounds. Finally, we required the hardest two subjets to be $b$-tagged with a tagging efficiency of 70\%. The mistag rate of the light jets faking as $b$-jets is taken to be a flat 2\%.

Having witnessed the prowess of a multivariate analysis (MVA) in Ref.~\cite{Banerjee:2018bio}, we refrain from doing the cut-based analysis (CBA) in this work~\footnote{Details of the CBA can be found in Ref.~\cite{Banerjee:2018bio}.}. First we construct fatjets with a cone radius of $R=1.2$, $p_T > 80$ GeV and $|y| < 2.5$ in the \texttt{FastJet}~\cite{Cacciari:2011ma} framework. Furthermore, we isolate the leptons with $p_T > 20$ GeV and $|y| < 2.5$ ($e\mu$) upon requiring that the total hadronic activity around a cone of radius $R=0.3$ about the lepton should be less than 10\% of its $p_T$. We select events with exactly to oppositely charged same flavour isolated leptons. Before performing the MVA, we select the final state with loose cuts on several variables, \textit{viz.}, $70 \; \text{GeV} < m_{\ell^+\ell^-} < 110$ GeV, $p_{T,\ell^+\ell^-} > 160$ GeV, $\Delta R_{\ell^+\ell^-} > 0.2$, $p_{T,\text{fatjet}} > 60$ GeV, $95 \; \text{GeV} < m_h < 155$ GeV, $\Delta R_{b_i,\ell_j} > 0.4$ and $\slashed{E}_T < 30$ GeV. The $\slashed{E}_T < 30$ GeV cut is imposed to almost completely remove the $t\bar{t}$ background. We also require that there is at least one fat jet associated with at least two $B$-meson tracks with $p_T > 15$ GeV. Furthermore, we require this fat jet to be double $b$-tagged. The $Z+$ jets, $gg \to Zh$, $t\bar{t}$ and $gg \to ZZ$ backgrounds being considerably subleading, the training of the boosted decision trees (BDT) is performed only with the SM $q\bar{q} \to Zh$ and $Zb\bar{b}$ samples upon using the following variables, \textit{viz.}, $p_T(\ell_1, \ell_2)$, $\Delta R(b_i \ell_j/\ell_1 \ell_2/b_1 b_2)$, where $i, j = 1,2$ and $b_i, b_j$ are the $b$-tagged subjets inside the fatjet, $m_{Z}$, $p_T(Z)$, $\Delta \phi(J,Z)$, $\slashed{E}_T$, $m_J$, $p_T(J)$, $p_T(b_1, b_2)$, $p_T(b_1)/p_T(b_2)$, $|y(J)|$, where $J$ is the reconstructed double $b$-tagged fatjet and $Z$ is the reconstructed $Z$-boson from the two isolated leptons. Our final variables of interest being the invariant mass of the reconstructed $Zh$-system and the three angles mentioned below, we do not consider these variables while training our samples. We utilise the \texttt{TMVA}~\cite{2007physics3039H} framework to train the signal and background samples and ensure that there is no overtraining~\cite{KS}.

\bibliography{references}    

\end{document}